# Gesture Based Interaction NUI: An Overview

Dr. Manju Kaushik*, Rashmi Jain**

*Associate Professor, Computer Science and Engineering, JECRC University, Jaipur, India*
**Research Scholar, Computer Science and Engineering, JECRC University, Jaipur, India*

*Abstract*- Touch, face, voice-recognition and movement sensors – all are part of an emerging field of computing often called natural user interface, or NUI. Interacting with technology in these humanistic ways is no longer limited to high-tech secret agents. Gesture recognition is the process by which gestures formed by a user are made known to the system. In completely immersive VR environments, the keyboard is generally not included, Technology incorporates face, voice, gesture, and object recognition to give users a variety of ways to interact with the console, all without needing a controller. This paper focuses on the emerging way of human computer interaction, Gesture recognition concept and gesture types.

*Keywords*- Natural User Interface, Gestures Recognition, Human Computer Interaction.

## I. INTRODUCTION

Buxton says everyone can enjoy using technology in ways that are more adaptive to the person, location, task, social context and mood. The rethinking of the way of interaction with computer system has paved the path of new challenges in the existing HCI techniques. For example, the most popular mode of HCI is based on simple mechanical devices –Keyboard and mice. These devices have grown to be familiar but inherently limit the speed and naturalness with which we can interact with computer [1].This limitation has become even more apparent with the emergence of novel display technology such as Virtual reality [2], [3], [4] .Thus there has been numerous research on providing quick technological solutions to complex problems of human interaction with computers. In the past interaction is based on punched cards, reserved to experts, than the interaction has evolved to the graphical interface paradigm. The interaction consists of the direct manipulation of graphic objects such as icons and windows using a pointing device. Even if the invention of keyboard and mouse is a great progress, there are still situations in which these devices are incompatible for HCI. This is particularly the case for the interaction with 3D objects. The 2 degrees of freedom (DOFs) of the mouse cannot properly emulate the 3 dimensions of space. The use of hand gestures provides an attractive and natural alternative to these cumbersome interface devices for human computer interaction [5].Gesture recognition is an approach towards an interactive, intelligent computing, efficient human computer interaction. It is the process by which the gestures made by the user are recognized by the receiver. Togive a specific definition to gestures is very difficult due to wide range of its applications and hardly a single definition can only express its one or two application. But a precise and specific definition is yet to be defined.

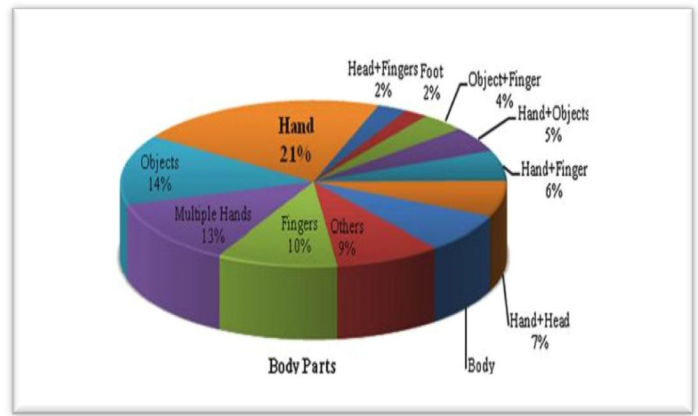

**Fig. 1** The*graph*shows the different body parts or objects identified in the literature employed for gesturing(Karam 2006)

Gestures are the motion of the body which is used with an intention to communicate with others. And to make this communication effective and successful both sender and receiver must have the same set of information for a particular gesture. A gesture is scientifically categorized into two distinctive categories: dynamic and static [1]. A dynamic gesture changes over a period of time while static gesture is observed at the spurt of time. A waving hand is an example of dynamic gesture and the stop sign is an example of static gesture.

Gestures are a body motion which expresses meaning through movement of body parts such as fingers, arms, hands, head and face. These movements 1] convey meaningful information or 2] interacting with the outer environment. They constitute one interesting small subspace of possible human motion. A gesture may also be perceived by the environment as a compression technique for the information to be transmitted elsewhere and subsequently reconstructed by the receiver. Gesture recognition has wide-ranging applications [6] such as the following:

- developing aids for the hearing impaired;
- enabling very young children to interact with computers;
- designing techniques for forensic identification;
- recognizing sign language;

- medically monitoring patients' emotional states or stress levels;
- lie detection;
- navigating and/or manipulating in virtual environments;
- communicating in video conferencing;
- distance learning/tele-teaching assistance;
- Monitoring automobile drivers' alertness/drowsiness levels, etc.

## II. GESTURE RECOGNITION

Gesture recognition is a process of recognizing and interpreting all the set of input data of static and dynamic gestures over a period of time. Gestures can be static (the user assumes a certain pose or configuration) or dynamic (with pre-stroke, stroke, and post-stroke phases). Some gestures also have both static and dynamic elements, as in sign languages. Again, the automatic recognition of natural continuous gestures requires their temporal segmentation. Often one needs to specify the start and end points of a gesture in terms of the frames of movement, both in time and in space. Sometimes a gesture is also affected by the context of preceding as well as following gestures. Moreover, gestures are often language-and culture-specific. They can broadly be of the following types [7].

- A. *hand and arm gestures:* recognition of hand poses, sign languages, and entertainment applications (allowing children to play and interact in virtual environments);

- B. *head and face gestures*: examples are: a) saying yes/no by nodding or shaking of head; b) showing direction by eye gaze; c) raising the eyebrows; d) opening the mouth to speak; e) winking, f) flaring the nostrils; and g) making expression by looks of surprise, happiness, disgust, fear, anger, sadness, contempt, etc.;

*Hand and arm gestures*

Human gestures typically constitute a space of motion expressed by the body, face, and/or hands. Of these, hand gestures are often the most expressive and the most frequently used. This involves: 1) a posture: static finger configuration without hand movement and 2) a gesture: dynamic hand movement, with or without finger motion.
Gestures may be categorized as given in the following list, such that as we precede downward this list, their association with speech declines, language properties increase, spontaneity decreases, and social regulation increases [7]**:**
- *Gesticulation*: spontaneous movement of hands and arms, accompanying speech. These spontaneous movements constitute around 90% of human gestures. People gesticulate when they are on telephone, and even blind people regularly gesture when speaking to one another;
- *Language like gestures*: gesticulation integrated into a spoken utterance, replacing a particular spoken word or phrase;
- *Pantomimes*: gestures depicting objects or actions, with or without accompanying speech;
- *Emblems*: familiar signs such as "V for victory," or other culture-specific "rude" gestures;
- S*ign languages: well-*defined linguistic systems. These carry the most semantic meaning and are more systematic, thereby being easier to model in a virtual environment. Hand gesture recognition consists of *gesture spotting* that implies determining the start and end points of a meaningful gesture pattern from a continuous stream of input signals and, subsequently, segmenting the relevant gesture.

The HCI interpretation of gestures requires that dynamic and/or static configurations of the human hand, arm, and even other parts of the human body, be measurable by the machine. First attempts to solve this problem resulted in mechanical devices that directly measure hand and/or arm joint angles and spatial position. This group is best represented by the so-called *glove-based devices* [8], [9], [10], [11], [12].

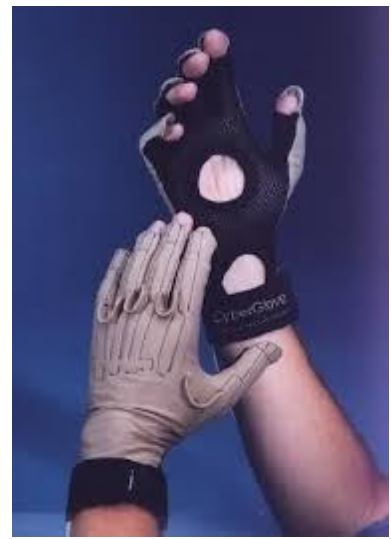

Fig:2 Glove based gestural interaction

Glove-based gestural interfaces require the user to wear a cumbersome device, and generally carry a load of cables that connect the device to a computer. This hinders the ease and naturalness with which the user can interact with the computer controlled environment. Even though the use of such specific devices may be justified by a highly specialized application domain, for example simulation of surgery in a virtual reality environment, the "everyday" user will certainly be deterred by such cumbersome interface tools. This has spawned active research toward more "natural" HCI techniques. To exploit the use of gestures in HCI it is necessary to provide the means by which they can be interpreted by computers. A new design tool lets people create 3D objects with their bare hands using a depth-sensing camera and advanced software algorithms that interpret hand movements. The Microsoft Kinect camera, which senses three-dimensional space. The camera is found in consumer electronics games that can track a person's body without using handheld electronics. Researchers created advanced algorithms that recognize the hand, understand that the hand is interacting with the shape and then modify the shape in response to the hand interaction.

*Face & Head Gesture*

Face is a unique feature of a human being. Humans can detect and identify faces in a scene with little or no effort. Their robustness is tremendous, considering the large changes inherent in the visual stimulus due to: 1) viewing conditions (such as variation in luminance); 2) facial expression; 3) aging; 4) gender; 5) occlusion; or 6) distractions such as glasses, hair style or other disguise. Human faces are non-rigid objects with a high degree of variability in size, shape, color, and texture. The goal of face detection is to efficiently identify and locate human faces regardless of their positions, scales, orientations, poses, and illumination. Any automated system for face and facial gesture recognition will have immense potential in criminal identification, surveillance, missing children retrieval, office security, credit card verification, video document retrieval, telecommunication, high-definition television (HDTV), medicine, human–computer interfaces, multimedia facial queries, and low-bandwidth transmission of facial data [13][14]. Recent advances in computer vision have led to efficient and robust head pose tracking systems, which can return the position and orientation of a user's head through automatic passive observation. Efficient methods for recognition of head gestures using discriminatively trained statistical classifiers have been proposed. In this work we use a robust real-time head tracking and gesture recognition system which was originally developed for interacting with conversational robots or animated characters [15]. There are two major approaches to automated face recognition [16].

- *Analytic:* Here flexible mathematical models are developed to incorporate face deformation and illumination changes. Discrete local (geometrical) features, such as irises and nostrils, are extracted for retrieving and identifying faces. The position of these features, with respect to one another, determines the overall location of the face. Standard statistical pattern recognition techniques such as HMMs [17], may be applied on these measurements. Other approaches include active contour models (Snakes) [18], wavelets [19], and knowledge- or rule-based techniques such as facial action coding system (FACS) [20], [21].

- *Holistic:* this involves gray-level template matching using global recognition. Here a feature vector is used to represent the entire face template. This approach includes ANNs [20]–[22], [23], [24], linear discriminates, PCA, singular value decomposition (SVD) using Eigen faces [25], [21], and optical flow [26], [27]–[29].

Research indicates that a fusion of both approaches often produces better (more stable) results, given an observation sequence, as compared to either approach alone [30]

### III. CONCLUSION

The importance of gestures recognition is a step further in human computer interaction, from touch to touch less interaction. Its application range from sign language recognition through medical rehabilitation to virtual reality. In this article researcher have provided an overview on a touch less interaction i.e. gesture recognition, with particular emphasis on hand gestures and facial expressions. Future technologies and research in human-computer interaction indicate that touch interaction and mouse input will not be the only broadly accepted ways users will engage with interfaces in the future. The future will also be touch less. Touch less interfaces (gestural) Natural Language Selection (voice recognition), GPS (proximity), and recommendation engines (data/insight/pattern matching) add convenience and a more natural cognitive relationship to information systems but require additional sensory connections making a user's experience that much more immersive, and interactive, in nature. Users are becoming remote controls where their minds and bodies control interactions in ways that were never before possible.